# The crystal structure, chemical bonding, and magnetic properties of the intercalation compounds $Cr_xZrTe_2$ (x = 0 - 0.3)

A.S. Shkvarin[1], A.A. Titov[1], A.I. Merentsov[1], E.G. Shkvarina[1], M.S. Postnikov[1], I. Píš[2,3], S. Nappini[3], P.A. Agzamova[1], A.S. Volegov[1,4], A.N. Titov[1],

[1]*Institute of Metal Physics, Russian Academy of Sciences-Ural Division, 620990 Yekaterinburg, Russia*

[2]*Elettra - Sincrotrone Trieste S.C.p.A., AREA Science Park Basovizza, S.S. 14 km 163.5, 34149 Trieste, Italy*

[3]*IOM-CNR, Laboratorio TASC, AREA Science Park Basovizza, S.S. 14-km 163.5, 34149 Trieste, Italy*

[4]*Ural Federal University, 620089 Yekaterinburg, Russia*

## Abstract

New intercalation compounds $Cr_xZrTe_2$ were synthesized in the Cr concentration range of x = 0 – 0.3. A thorough study of the crystal and electronic structure has been performed. It was found that there is competition in the distribution of the Cr atoms over the octa- and tetrahedral sites in the van der Waals gap, depending on the Cr content. The ordering of the Cr atoms was found at x = 0.25; at the same time, the lattice symmetry decreases from trigonal P-3m1 to monoclinic F2/m. This ordering stabilizes the octahedral coordination of the Cr atoms by Te atoms. The analysis of the experimental data on the electronic structure and DOS calculations showed that the Cr 3d states are spin-split.

## Introduction

Layered transition metal dichalcogenides (LTMDs) and their intercalation compounds have attracted attention for several decades [1–9]. These materials demonstrate superconductivity [8,10–12], a state with a charge density wave [13–16]. These materials are suggested to be good candidates to use as cathode materials for lithium electrochemical power sources [17–20] and photodetectors [21–23]. The use of LTMDs in spintronics [24–28] is one of the promising areas of modern material science. Two important features have been revealed from the study of the electronic structure of the $Cr-TiCh_2$ (Ch = Se, Te) systems [29–31]: 1) the Cr states are located directly at the Fermi level or near to it; 2) these states are spin-polarized with a nearly 1 eV splitting of the spin subbands. However, the formation of a 100 % spin polarization at the Fermi level is complicated by the fact that there is no gap between the Ti 3d-derived conduction band and the Se/Te np-derived valence band in $TiSe_2$ and $TiTe_2$. This leads to the equality of the energy of the Cr spin-polarized states and the Ti and Ch (Ch = Se, Te) spin-unpolarized states. In its turn, $TiS_2$ is a narrow-gap semiconductor, however, it is known for its non-stoichiometry, because it is always contaminated with an excess of Ti, which provides a significant concentration of excess electrons [1]. The LTMDs free of transition metal excess but keeping a bandgap at the Fermi level are zirconium dichalcogenides $ZrSe_2$ and $ZrTe_2$ (the bandgap for $ZrSe_2$ is nearly 1 eV [32], for $ZrTe_2$ there is a pseudogap [33]). The intercalation of chromium in such lattices can provide a complete spin polarization of states at the Fermi level, but only in the case of the Cr 3d states will be located directly at the Fermi level. The current study is aimed to the synthesis of the $Cr_xZrTe_2$ intercalation system and the study of its crystal and electronic structure to determine the position of the Cr states relative to the Fermi level.

## Experimental

Polycrystalline samples of the $Cr_xZrTe_2$ system were synthesized with Cr concentration in the range of $0 \leq x \leq 0.3$ from preliminary prepared $ZrTe_2$ and metallic chromium (purity 99,93%). $ZrTe_2$ was obtained from the starting elements: Zr (iodine purification, 99.95%) (VSMPO-AVISMA) and Te (after double distillation in vacuum, RPA, 99.99%). The required quantity of elements was sintered at a temperature of 1150 °C for 7 days in sealed quartz ampoules evacuated to $10^{-5}$ torr. Then, the ampoules were opened, the samples were grinded, pressed, again sealed in evacuated quartz ampoules, and annealed for homogenization under the same conditions. $Cr_xZrTe_2$ samples were obtained by standard thermal intercalation [34]. For this, the corresponding amount of Cr (99.9%) (как правильно писать наш хром?) and preliminary prepared $ZrTe_2$ powder was mixed in the required proportion, pressed into pellets, sealed in quartz ampoules evacuated to $10^{-5}$ Torr and sintered at 1100 °C for a week. After this, the resulting product was thoroughly grinded in an agate mortar, pressed and annealed again under the same conditions for complete homogenization. $Cr_{0.25}ZrTe_2$ and $Cr_{0.3}ZrTe_2$ samples were additionally annealed at 750 °C for a week. The final product was macroscopically homogeneous. The structure and phase composition of the samples were studied at room temperature using X-ray powder diffraction on an XRD 7000 Maxima Shimadzu diffractometer (Cu Kα radiation, graphite monochromator, with a 2θ range of 5 - 90° and a step size of 0.02°) in the Collective Use Center 'Ural-M' of the Institute of Metallurgy, Russian Academy of Sciences, Ural Division.

The crystal structure refinement was performed using GSAS (General Structure Analysis System) [35]. For all samples, the following model of the crystal structure was used: P-3m1 space group (164) [36]. The used unit cell consisted of the following atoms: Zr(1a) (0 0 0), Te(2d) (⅓ ⅔ z), Cr(1b) (0 0 ½) - octahedral sites, Cr(2d) (⅓ ⅔ z) - tetrahedral sites [36]. Fraction values were constrained by total Cr concentration taking into account the multiplicity of the positions. A pseudo-Voigt function with a Finger–Cox–Jephcoat peak asymmetry correction (peak profile type No. 3) was used for the profile fitting. The background was approximated by Chebyshev polynomial with 20 coefficients. The atomic coordinates of the Te and Cr atoms in the tetrahedral sites along with the occupation of the octahedral and tetrahedral sites by the Cr atoms were refined.

The study of the electronic structure by soft X-rays spectral methods requires a fresh clean surface. Such a surface can be obtained by cleaving a sufficiently large single crystal in ultra-high vacuum conditions. Therefore, the condition for the successful completion of the study of the electronic structure is the growth of single crystals of a sufficiently high quality and size. $Cr_xZrTe_2$ single crystals with a chromium content of x = 0.25 were grown using gas transport reactions technique in evacuated quartz ampoules placed in the temperature gradient of nearly 5 °C/cm with iodine as a carrier gas [37]. The crystals were grown on the cold end of the ampoule. The crystals were in the form of thin plates with a size of approximately $3\times3\times0.005$ mm$^3$.

The X-ray photoelectron (XPS), X-ray absorption (XAS), and resonant X-ray photoelectron (ResPES) spectra of $Cr_{0.25}ZrTe_2$ were obtained at room temperature (RT) the BACH beamline of CNR [38] at the Elettra synchrotron facility (Trieste, Italy). All of the samples were cleaved *in situ* in a vacuum chamber at a pressure lower than $1\times10^{-9}$ Torr. The purity of the surface was confirmed by the absence of oxygen and carbon peaks in the XPS survey spectra. The XPS spectra were collected using 600 eV photon energy and a hemispherical electron energy analyzer (VG Scienta, model R3000) at an angle of 60° with respect to the X-ray incidence beam direction. The spectra were recorded at normal emission angle. Binding energies were calibrated to Au $4f_{7/2}$ (84.0 eV) peak measured on a gold reference. The total instrumental resolution was 150 meV. Zr $M_{2,3}$ and Cr $L_{2,3}$ XAS were measured in total electron yield with linearly polarized light at an incidence angle of 30°. The photon energy resolution was set to 80 meV and the energy scale was calibrated with an accuracy of 0.1 eV using Au 4f XPS spectra measured with the fundamental and second harmonic of the monochromator.

First principles calculations have been performed to obtain a theoretical description of the density of states of the studied materials near the Fermi level. The total density of state (TDOS) and partial density of states (PDOS) calculations were performed using the Vienna Ab initio Simulation Package (VASP) [39,40]. We used the projector augmented-wave (PAW) method [41] with the Perdew-Burke-Ernzerhof (PBE) type of exchange-correlation functional within the general gradient approximation (GGA) [42]. The cut-off energy was chosen to be $E_{cutoff}$ = 350 eV and 18×18×1 Monkhorst-Pack grid of k-points was used during calculations. The self-consistent calculation was finished when the total energy change became less than 0.0001 eV. Experimentally determined parameters of the crystal structure were used in these calculations. The "Uran" supercomputer of IMM UB RAS was used for these calculations.

The temperature and magnetic field dependences of the magnetic moment of the samples were measured using the MPMS XL 7 EC setup (Quantum Design, USA). Samples were glued on Kapton tape and then were placed in a plastic tube. Kapton tape was coaxial to the tube; the sample was located in the middle of the tube. The length of the tube and Kapton was equal to 15 cm, 5 times longer than the length of the gradient meter. This sample location allows to avoid the parasitic contribution of the sample holder of another shape. The error in measuring the magnetic moment does not exceed 5%. Temperature dependences were measured with the heating rate of 2 K/min. Magnetic field dependencies were measured in Hysteresis Mode.

## Results and discussion

### Crystal structure

The Cr concentration dependences of the *a* and *c* lattice parameters are shown in Fig. 1. At low Cr concentration (x ≤ 0.1), the sites in the van der Waals gap, which are octahedrally coordinated by the Te atoms (octahedral sites), are occupied at first. An increase in the occupation of these sites leads to a compression of the unit cell (Fig. 1b). In general, this can be caused by the compression of both the Te-Zr-Te sandwich and interlayer space. The first option is realized for $Cr_xZrTe_2$ (see the constancy of the van der Waals gap width, Fig. 1c). The compression of the Te-Zr-Te sandwiches may indicate the effective charge transfer between the Zr, Cr, and Te sublattices and, therefore, changes in chemical bond lengths.

An increase in the Cr content from (0.1 < x ≤ 0.15) leads to a distribution of 5 at. % of the Cr atoms over the sites in the van der Waals gap, which are octahedrally coordinated by the Te atoms (tetrahedral sites). A further increase in the Cr content (x = 0.2) leads to the displacement of the Cr atoms from the octahedral to tetrahedral sites. This redistribution of the Cr atoms interrupts the lattice contraction and makes the *c* lattice parameter independent of the Cr content at 0.15 ≤ x ≤ 0.2. Surprisingly, at x = 0.25 all the Cr atoms occupy only the octahedral sites again. It becomes clear taking into account that for the isostructural compounds $M_{0.25}TiSe_2$ (M = Fe, Co, Ni) [43] the ordering of the intercalated atoms M has been observed; this ordering is possible only in the case of all the Cr atoms occupy the octahedral sites. However, we observed a similar ordering for the $Cr_{0.25}ZrTe_2$ sample annealed at 750 °C, whereas the data listed in Table 1 are related to the samples obtained at 1150 °C without additional annealing.

The compression of the lattice in the direction normal to the basal plane is typical for layered titanium dichalcogenides intercalated with transition metals [30,43–47], but in all of these cases, the intercalated metal atoms occupy only the octahedral sites in the interlayer space. Therefore, there is no data about the intercalant concentration dependence of the lattice parameters in the case of the occupation of tetrahedral sites. The exception is the intercalation systems with alkali metals [48–50], in which an increase in the concentration of the intercalated atoms leads to an increase in the $c_0$ parameter at the occupation of both octahedral and tetrahedral sites.

The occupation of the tetrahedral sites by the intercalated atoms is observed for the systems based on $ZrCh_2$ (Ch = S, Se, Te), as, for example, for $Fe_xZrCh_2$ [51,52]. In another work [53] aimed to the synthesis and study of the $M_xZrS_2$ systems (M = Fe, Co, Ni,) it was shown that the Fe atoms occupy both the octahedral and tetrahedral sites; the occupation of the octahedral sites was much higher than the tetrahedral ones. In the $M_xZrS_2$ (M = Fe, Co, Ni) systems, the structural analysis was performed only for $Ni_{0.5}ZrS_2$, in which the Ni atoms occupied the tetrahedral sites. Iron intercalation into $ZrS_2$ led to an increase in the $c$ parameter, whereas the intercalation of the Co and Ni atoms into $ZrS_2$ led to the lattice contraction in the direction normal to the basal plane. This was due to the localization of electrons introduced from intercalated atoms and change in the Zr charge state from $Zr^{4+}$ to $Zr^{3+}$. Thus, the available data on the influence of the occupation of the tetrahedral sites on the compression or expansion of the lattice in the $c$ axis direction remain contradictory.

In the $M_xTiCh_2$ intercalation systems (M is a transition metal), the lattice compression in the $c$ axis direction was explained by the formation of a covalent bond of the intercalated atoms with Ti atoms due to the Ti $3d_z^2$/M $3d_z^2$ hybridization. The occupation of the tetrahedral sites by the intercalated atoms does not allow them to form these hybrid states. Therefore, it can be expected that in the case of the occupation of the tetrahedral sites by the intercalated atoms, the lattice compression will not be observed. Probably, contrariwise, the isotropic expansion of it will be observed.

Thus, in $Cr_{0.2}ZrTe_2$, in addition to the disappearance of the contracting factor (in the case of the Cr atoms occupy the octahedral sites, the Zr $4d_z^2$/Cr $3d_z^2$ hybridization leads to the contraction of the lattice in the c axis direction), the occupation of the tetrahedral sites with a high value of "z" coordinate $z_{Cr} = 0.71(1)$ leads to an expansion of the interlayer space.

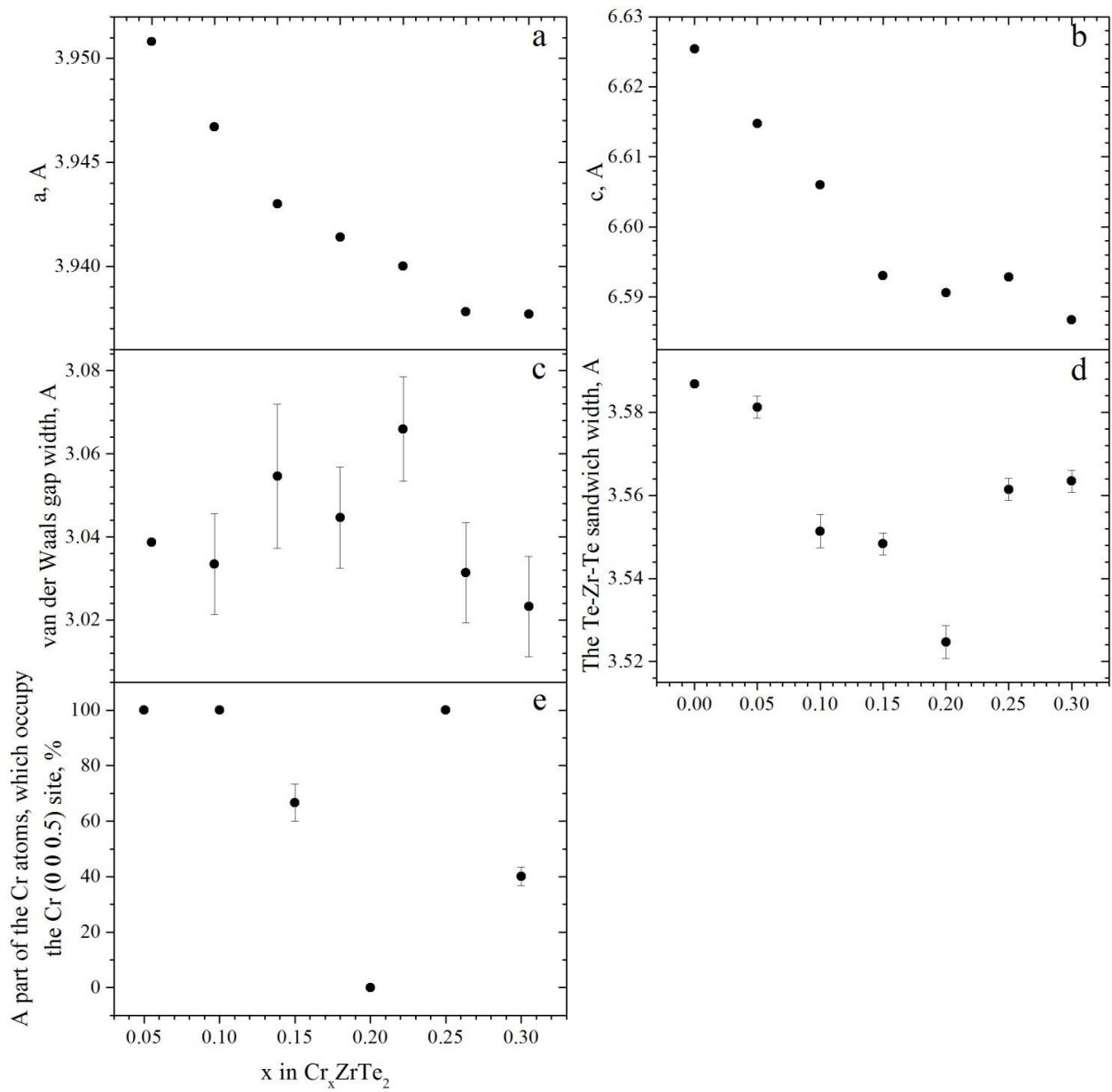

**Figure 1.** The concentration dependencies of: a) the *a* lattice parameter; b) the *c* lattice parameter c) the van der Waals gap width; d) the "sandwich" width; e) the percentage of Cr atoms located in the octahedral sites (100% corresponds to the total Cr amount in the sample).

**Table 1.** The refinement results for the intercalation $Cr_xZrTe_2$ (x= 0 - 0.3) system.

| x | A | c | Te (⅓ ⅔ z) | Cr (0 0 ½), multiplicity = 1 | | Cr (⅓ ⅔ z), multiplicity = 2 | | $R_F^2$, % | wRp, % | Rp, % | $\chi^2$ |
|---|---|---|---|---|---|---|---|---|---|---|---|
| | | | z | | occupation | z | occupation | | | | |
| 0.05 | 3.9467(1) | 6.6147(1) | 0.2707(2) | | 0.05 | - | - | 12.27 | 11.32 | 8.27 | 1.769 |
| 0.1 | 3.9430(1) | 6.6060(2) | 0.2688(3) | | 0.1 | - | - | 8.83 | 12.30 | 9.45 | 2.125 |
| 0.15 | 3.9414(1) | 6.5930(1) | 0.2691(2) | | 0.10(1) | 0.58(2) | 0.026(5) | 8.50 | 13.86 | 10.60 | 2.240 |
| 0.2 | 3.9400(1) | 6.5906(1) | 0.2674(3) | | - | 0.71(1) | 0.1 | 13.57 | 11.58 | 7.92 | 2.097 |
| 0.25 | 3.9378(1) | 6.5928(1) | 0.2701(2) | | 0.25 | - | - | 9.78 | 14.20 | 10.89 | 2.360 |
| 0.3 | 3.9377(1) | 6.5867(1) | 0.2705(2) | | 0.12(1) | 0.510(7) | 0.09(1) | 12.14 | 15.78 | 11.90 | 2.685 |

The question about the reason for the displacement of the Cr atoms between octa- and tetrahedral sites arises. We can answer this question analyzing the diffraction patterns for $Cr_{0.25}ZrTe_2$ and $Cr_{0.3}ZrTe_2$ annealed at 750 °C (Fig. 2). The additional superstructure peaks (marked in Fig. 2 with blue arrows) corresponding to the $a_0 \times 2a_0\sqrt{3} \times 2c_0$ (here $a_0$, $c_0$ – lattice parameters for $Cr_{0.3}ZrTe_2$ without ordering) superstructure (space group F2/m) appear. The same ordering is formed in $Cr_{0.25}ZrTe_2$, however the superstructure peaks for $Cr_{0.3}ZrSe_2$ are most pronounced.

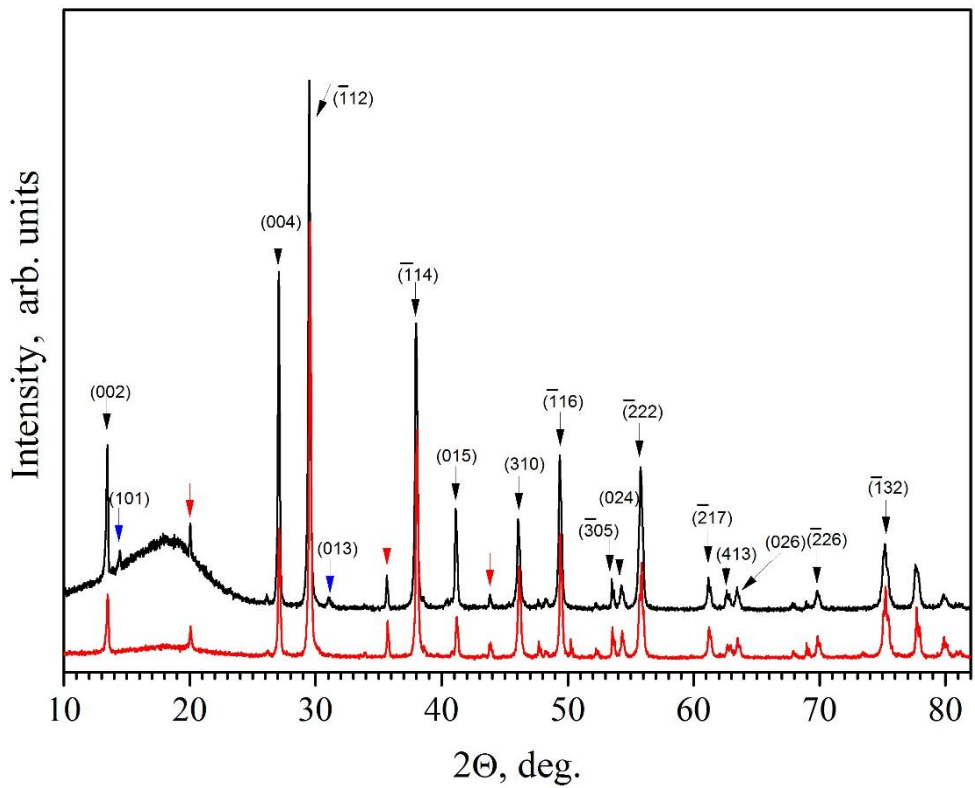

**Figure 2.** X-ray diffraction pattern for $Cr_{0.3}ZrTe_2$ annealed at 750 °C (black line) and sintered at 1150 °C (red line). The red arrows mark the zircon ($ZrSiO_4$) lines. The zircon phase is caused by the reaction of the small fragments of a quartz ampoule with zirconium during the annealing. The total amount of this phase does not exceed 5%. The blue arrows mark the most pronounced superstructure peaks.

This type of ordering was previously observed for the $M_xTiCh_2$ (Ch = S, Se, Te) intercalation compounds (M - 3d-transition metal) [43]. This ordering appears due to the interaction between the Ti-Ch-M-Ch-Ti covalent complexes, which appear due to the M $3d_z^2$/Ti $3d_z^2$/Ch np (n = 3, 4, 5 for Ch = S, Se, Te, respectively) hybridization. In the case of the $Fe_xTiSe_2$ system, we showed [54] that this bonding is accompanied by the energy gain of about 1.4 eV, so that the formation of a chemical bond between the complexes is energetically favorable. One can expect that in the $Cr_xZrTe_2$ system, the ordering associated with the interaction between the covalent complexes will be energetically favorable too. However, this ordering is observed only when the intercalated atoms occupy the octahedral sites. Probably, it is the reason that causes the displacement of the Cr atoms into the octahedral sites at x = 0.25. At low Cr concentration (x = 0.05 and 0.10), the Cr atoms can be considered as isolated ones from each other and, therefore, non-interacting. Therefore, the displacement of the Cr atoms from the octahedral to tetrahedral sites can be associated with the Coulomb repulsion between the Cr atoms.

## XPS

The core level spectra provide information about the chemical bonding in the studied material. The Ch nd (Ch = Se,Te; n =3, 4) spectra are very sensitive to the quality of the studied surface [55,56]. The Te 4d and Zr 3d spectra for $Cr_xZrTe_2$ are shown in Figure 3.

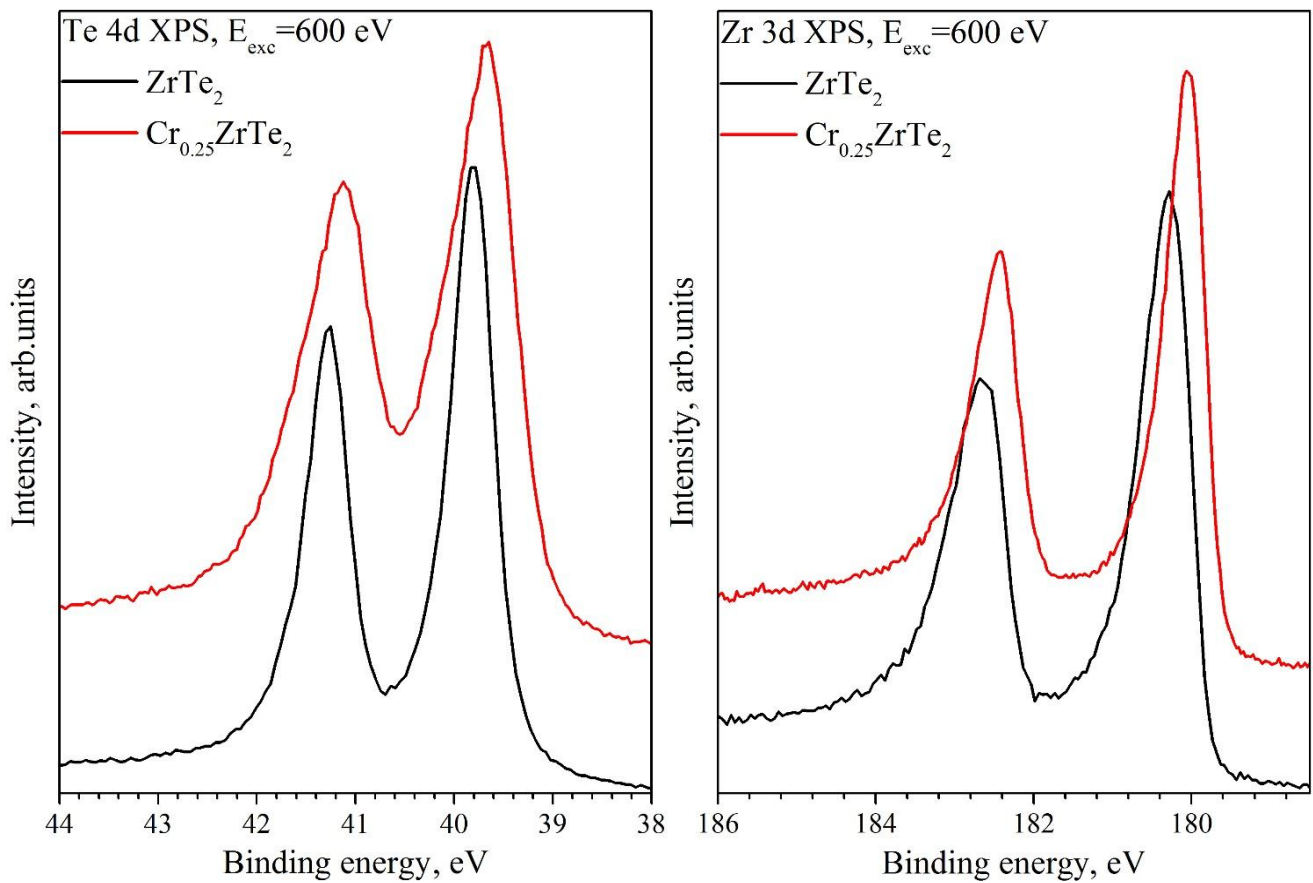

**Figure 3**. Te 4d and Zr 3d core level spectra for $ZrTe_2$ and $Cr_{0.25}ZrTe_2$.

The Te 4d core level spectra for $ZrTe_2$ and $Cr_{0.25}ZrTe_2$ are different. A 0.16 eV shift of the spectrum towards lower binding energies, as well as its significant broadening, are observed for the Te 4d spectrum for $Cr_{0.25}ZrTe_2$ with respect to $ZrTe_2$. The Zr 3d spectrum for $Cr_{0.25}ZrTe_2$ shifts on 0.2 eV towards lower binding energies with respect to $ZrTe_2$, but the line width does not change. This indicates, at first, a shift in the Fermi level, and, at second, a significant involvement of the Te states into the formation of the chemical bonds between the Cr atoms and the host lattice. This result is consistent with the previously reported data [57], which indicates that the Cr intercalation substantially changes the Zr-Te chemical bond.

The valence band (VB) spectra are shown in Fig. 4.

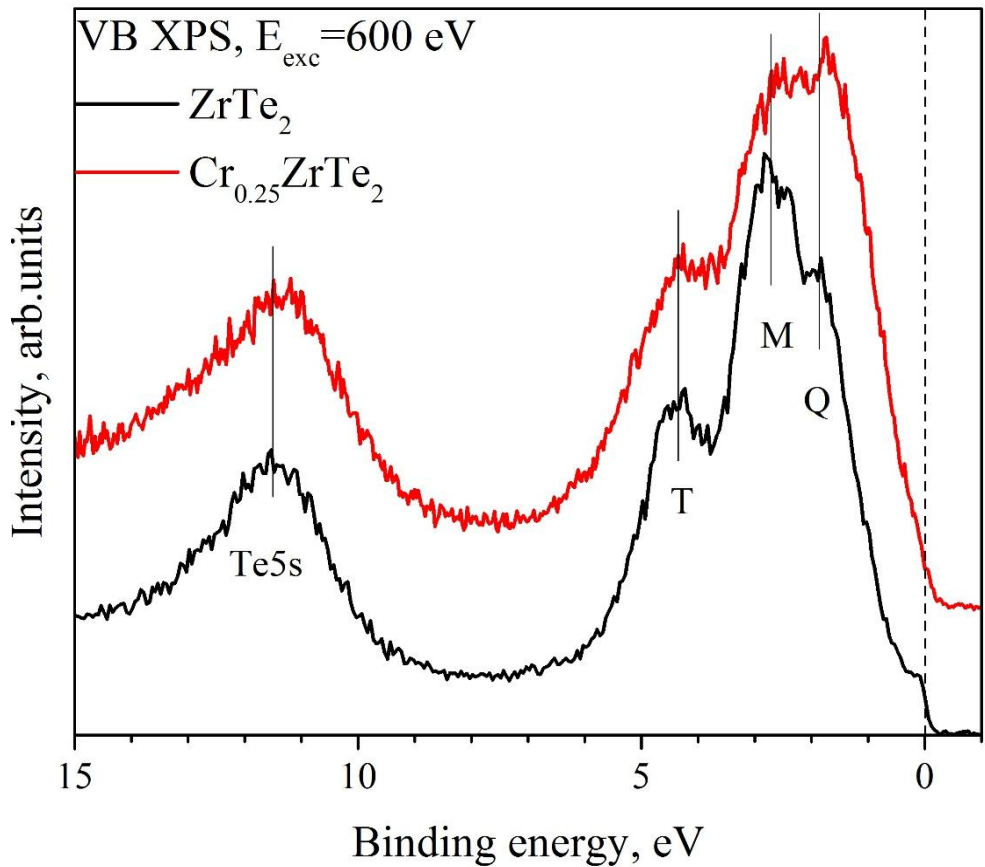

**Figure 4**. VB spectra for $ZrTe_2$ and $Cr_{0.25}ZrTe_2$.

The Cr intercalation into $ZrTe_2$ does not change the general structure of the VB spectrum; the spectra remain typical for most transition metal dichalcogenides [58,59]. Most changes were observed directly below the Fermi level. A change in the ratio of the intensities of the peaks M and Q was observed. The step at the Fermi level becomes blurred. This is probably due to the formation of new states near the Fermi level, but not directly at the Fermi level. A shift of the Te 5s line toward lower binding energies can be noted, as is observed for the other core level spectra.

## XAS

The shape of the absorption spectra depends on the local environment of the corresponding atom. The Cr $L_{2,3}$ and Zr $M_{2,3}$ XAS spectra are shown in Fig. 5.

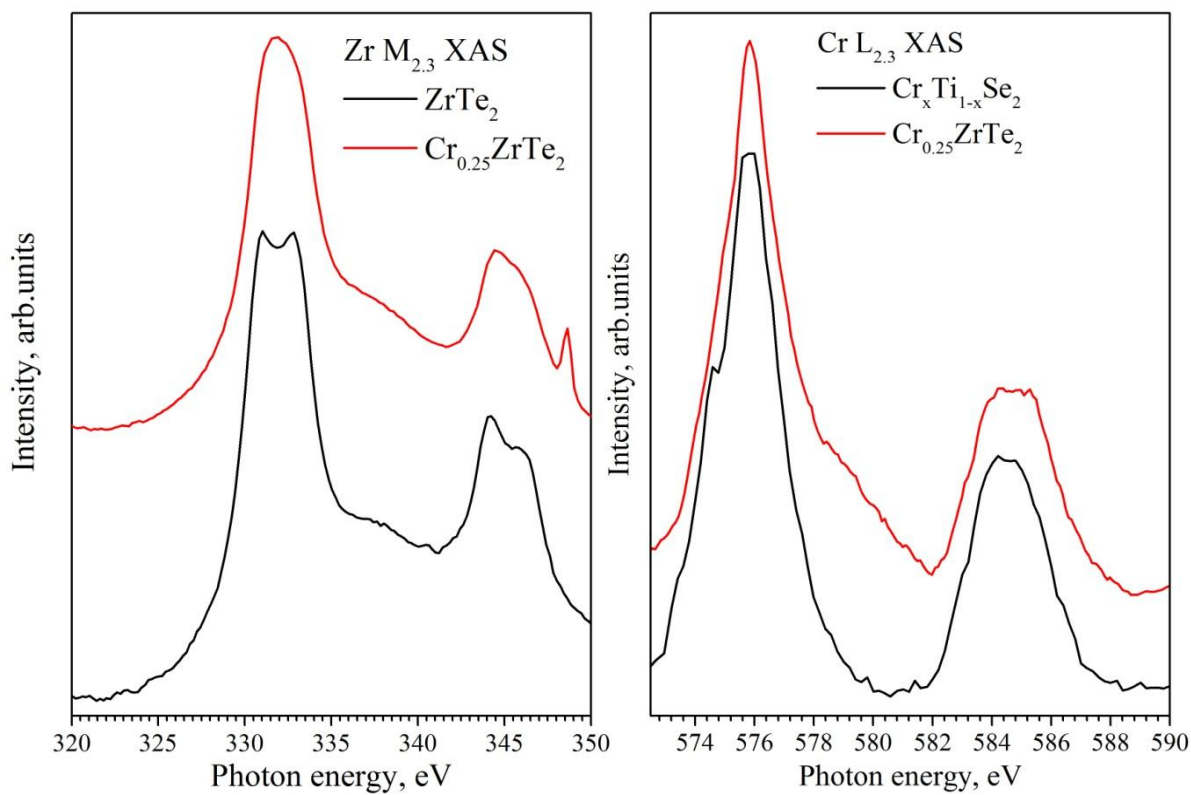

**Figure 5**. Zr $M_{2,3}$ XAS and Cr $L_{2,3}$ XAS for $ZrTe_2$ and $Cr_{0.25}ZrTe_2$.

One can see that Cr $L_{2,3}$ XAS for $Cr_{0.25}ZrTe_2$ is similar to that for the $Cr_xTi_{1-x}Se_2$ system [60], in which the Cr charge state is $Cr^{3+}$ one. Zr $M_{2,3}$ XAS for $Cr_{0.25}ZrTe_2$ significantly differs from that for $ZrTe_2$. The Cr intercalation into $ZrTe_2$ leads to the broadening of Zr $M_{2,3}$ XAS, however, Cr $L_{2,3}$ XAS does not broaden. This indicates a significant change in the interaction of the Zr atoms with the local environment, i.e. with the Te atoms, which can be explained by a significant distortion of the Te sublattice due to the Cr intercalation; these results are in a good agreement with those reported in [57].

## ResPES

The study of systems containing both the Te and Cr atoms by spectral methods is complicated by the fact that more intense Te lines overlap with less intense Cr lines. Therefore, it is impossible to obtain Cr core level spectra. The VB spectrum contains lines of all elements from the studied chemical composition. To distinguish the Cr states in the valence band, a ResPES experiment was performed.

Resonant photoelectron spectroscopy provides element-specific information about the energy position of electronic states. Photoemission of d electrons for 3d metals and their compounds is significantly enhanced when the incident photon energy slightly exceeds the binding energy of a selected level upon excitation of np electrons (n = 2; 3) to unfilled 3d states, this leads to resonant emission of electrons [61–64]. The result is a coherent process in which an np-electron from the initial state becomes excited to an empty 3d state with the formation of an intermediate bound state (np, 3d).

The VB spectra in the Cr 2p-3d resonant excitation mode are shown in Figure 6.

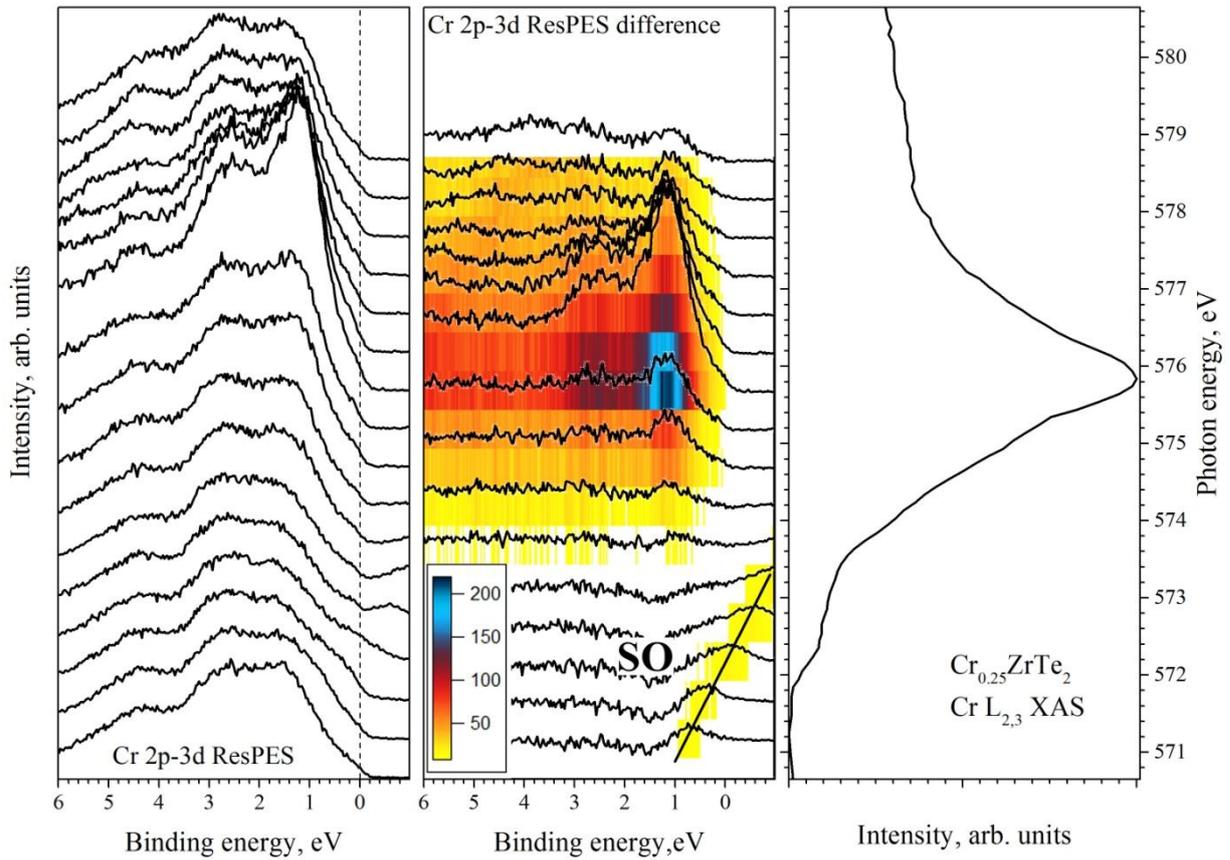

**Figure 6**. VB spectra obtained in the Cr 2p-3d resonant excitation mode for $Cr_{0.25}ZrTe_2$. The difference spectra, which were obtained by the method described in Ref. [31], are shown in the central panel in the form of an image plot. On the right panel: corresponding Cr $L_3$ XAS spectrum. The excitation energies for the VB spectra on the left and central panels correspond to the photon energies of Cr $L_3$ XAS from the right panel (for the lower VB spectrum $E_{exc}$ = 570.7 eV, for the upper VB spectrum $E_{exc}$ = 578.7 eV).

The second reflection order of the Te 3d state is marked as SO in the central panel, Fig. 6. The SO contribution to the resonant spectra is far from the main resonance; therefore, it does not affect the states below the Fermi level. A strong resonance band with a binding energy of 1.1 eV and a band with a lower intensity and binding energy of about 2.5 eV are seen in ResPES spectrum. Both the binding energy and the excitation energy position of these bands indicates the localized character of the Cr 3d electrons. We believe that these states are those additional states that Zhang et. al. has observed in [57] at the M point of the Brillouin zone slightly below the bottom of the conduction band. Although it should be noted, that in [57] the binding energy of these additional states was equal to 0.2 – 0.3 eV in contrast to 1 eV in our case. Since in [57] the chemical composition of the studied crystal was $Cr_{0.4}ZrTe_2$, this discrepancy is likely due to the dependence of the energy position of these states on the Cr concentration in $Cr_xZrTe_2$.

Calculations

The first-principle calculation was performed using the experimentally obtained crystal structure to simulate the total and partial density of states (TDOS and PDOS). The unit cell of $ZrTe_2$ (P-3m1 space group, No. 164 [36]) was doubled in X, Y, Z directions; the resulting supercell contained 8 cells of the starting $ZrTe_2$. The Cr atoms were uniformly distributed over the octahedral sites at the interlayer space. A fragment of the crystal structure is shown in Figure 7.

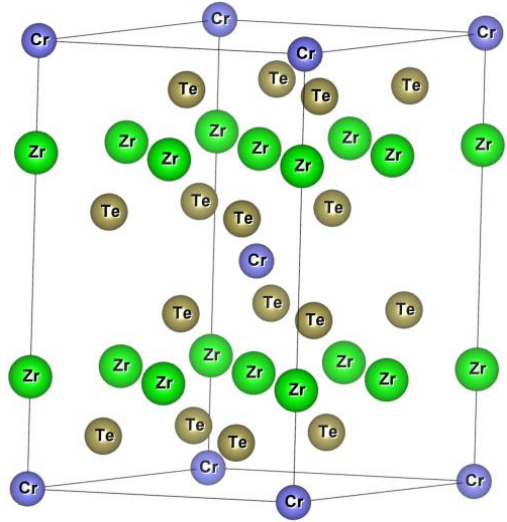

**Figure 7**. The supercell for $Cr_{0.25}ZrTe_2$ used for the DOS calculation.

TDOS calculated for $ZrTe_2$ and $Cr_{0.25}ZrTe_2$ together with the experimental VB spectrum is shown in Figure 8.

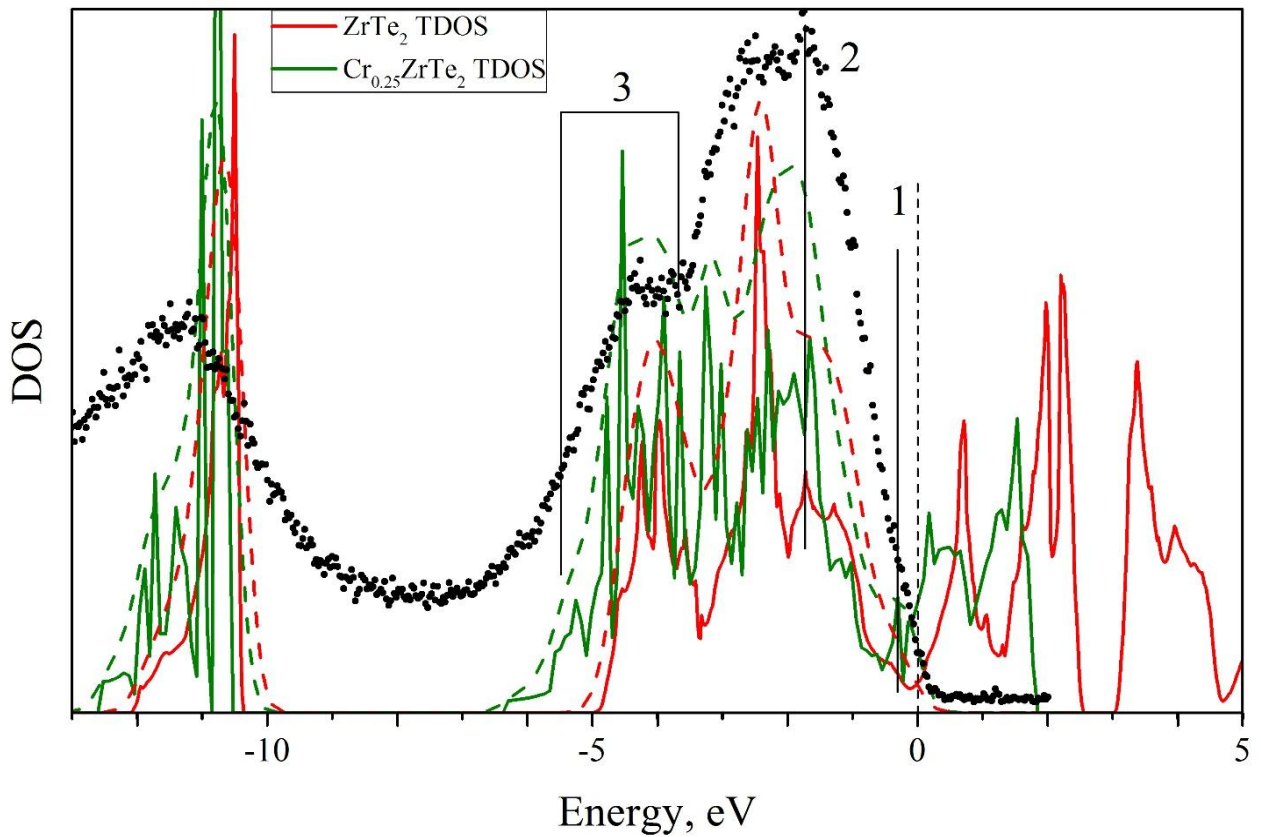

**Figure 8.** Calculated TDOS for $Cr_{0.25}ZrTe_2$ and $ZrTe_2$ (solid lines) along with experimental VB at $E_{exc}$ = 594 eV (dotted line). Calculated TDOS after Gaussian broadening are shown by dashed lines.

One can see that $Cr_{0.25}ZrTe_2$ TDOS is more uniform and wider than $ZrTe_2$ TDOS. Therefore, the VB spectrum for $Cr_{0.25}ZrTe_2$ is broader than that for $ZrTe_2$. One can distinguish several bands (marked with numbers in Fig. 8), which contribute the experimental VB spectra and make the VB spectrum for $Cr_{0.25}ZrTe_2$ different from that for $ZrTe_2$. Peak 2 in Fig. 8 corresponds to the Q band in Fig. 4, peak 3 in Fig. 8 corresponds to the T band in Fig. 4. Peak 1 in Fig. 8 causes the blurring of the step at the Fermi level. We believe that the DOS calculation well describes all the features in the experimental VB spectrum.

Fig. 9 shows Cr, Te, and Zr partial DOS for $Cr_{0.25}ZrTe_2$. Spin-polarized calculation reveal the spin splitting of the Cr states. Several bands for the Cr states can be distinguished: a band directly at the Fermi level (marked with vertical dashed line), a band at 1.1 eV and a band at 2.2 eV (marked with vertical solid lines). One can see that the band at the Fermi level is strongly hybridized with the Zr and Te states. On the contrary, the band with an energy of 1 eV is formed almost completely by the Cr states. It is exactly the band that is observed on the ResPES obtained in the Cr 2p-3d resonant excitation mode. The band with an energy of 2.2 eV also contributes to the VB spectrum obtained in the Cr 2p-3d resonant excitation mode (see Fig. 6).

.

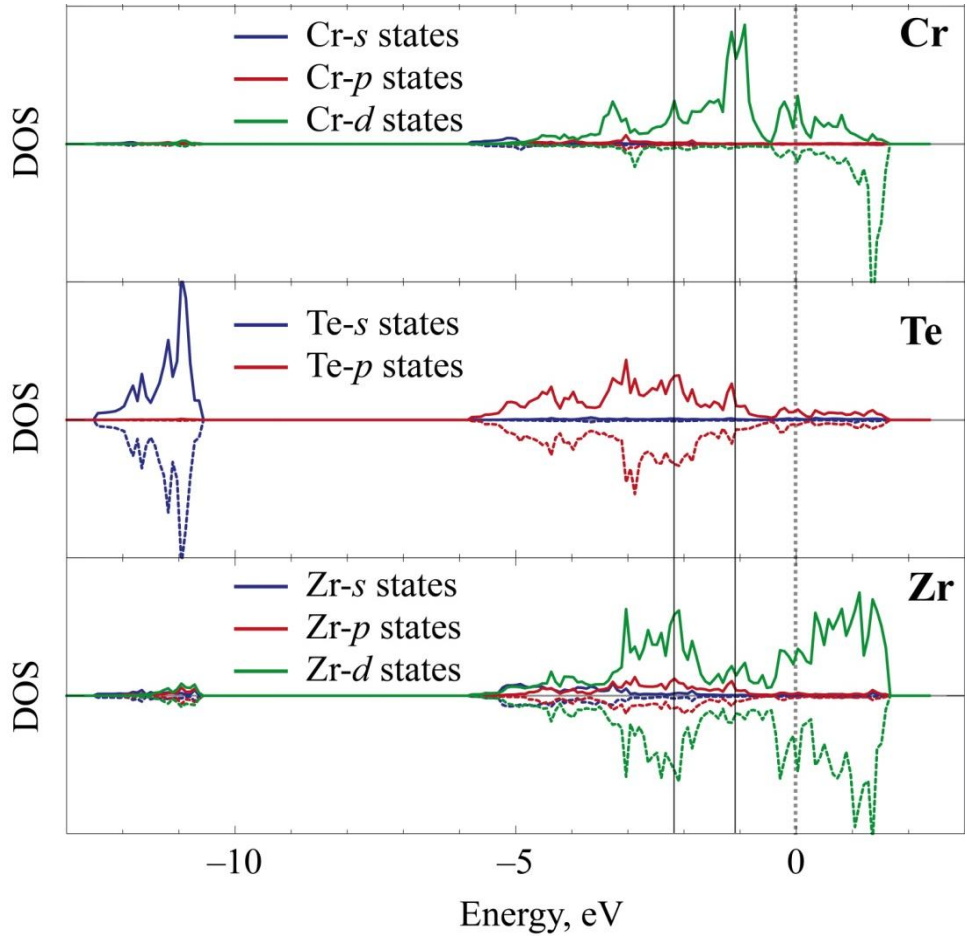

**Figure 9.** Cr, Te and Zr partial DOS (from top to bottom) for $Cr_{0.25}ZrTe_2$.

Magnetic properties

Fig. 10 shows the temperature dependences of the magnetic susceptibility for $Cr_xZrTe_2$ polycrystalline samples. One can see the spin-glass magnetic ordering for $Cr_xZrTe_2$ in the Cr concentration range of x = 0.05 - 0.3. An increase in the external magnetic field leads to a decrease in the magnetic susceptibility in the region of its maximum value.

The value of the effective magnetic moment $\mu_{eff}$, calculated from the temperature dependence of the magnetic susceptibility $\chi$ in the temperature range above the ordering temperature, is shown in Table 2. The obtained $\mu_{eff}$ values are close to that for the $Cr^{3+}$ spin moment (3.87 $\mu_B$). A slight increase in the value of $\mu_{eff}$ at x = 0.2 can be caused by a slight delocalization of Cr 3d electrons due to the absence of the Cr 3d/Zr 4d hybridization. Indeed, all the Cr atoms occupy the tetrahedral sites in $Cr_{0.2}ZrTe_2$, therefore the Cr 3d orbitals cannot overlap with the Zr 4d ones.

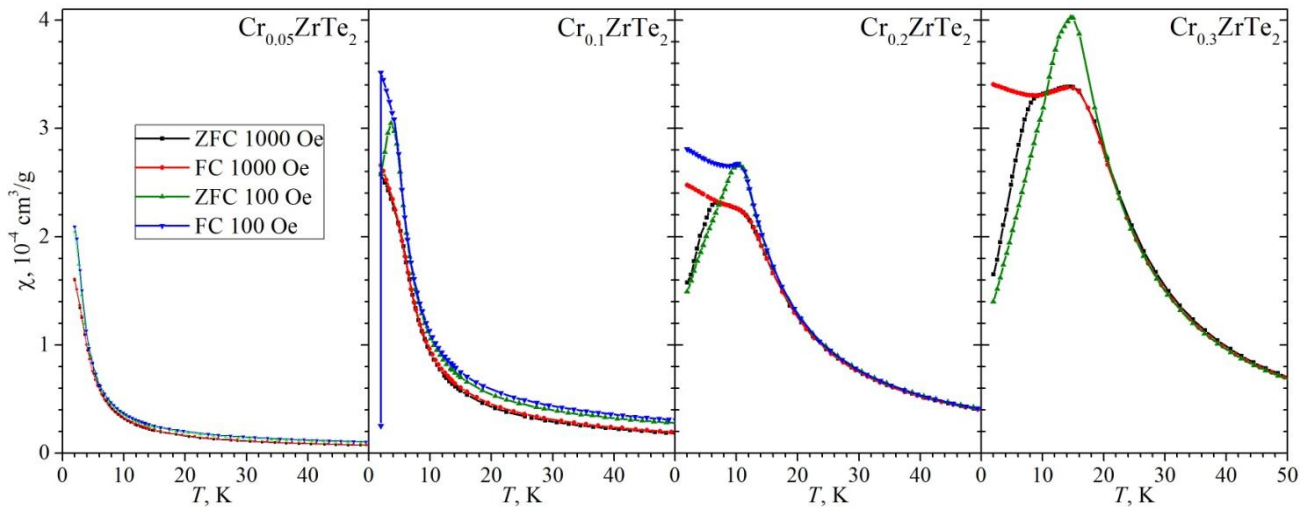

**Figure 10**. Temperature dependences of the magnetic susceptibility for $Cr_xZrTe_2$ near the temperature of the magnetic ordering measured in zero field cooling (ZFC) and field cooling (FC) modes. The measurements were performed in external magnetic fields of 100 Oe and 1000 Oe.

As can be seen from Table 2, the character of the magnetic interaction in the Cr sublattice is almost independent of either the Cr concentration or the external magnetic field. However, it is clearly seen that for all studied compounds the temperature of the magnetic ordering decreases with the increase of the external magnetic field. Since the temperature dependences of the magnetic susceptibility for $Cr_xZrTe_2$ passes through the maximum typical for the antiferromagnetic materials, but the common features of the ordering are typical for the spin (or cluster) glass state, a competition between ferro- and antiferromagnetic interactions can be expected. (не понял, откуда взялось про кластерное стекло) It is evident that an increase in the external magnetic field enhances the ferromagnetic contribution and weakens the antiferromagnetic one. This explains the decrease in the ordering temperature with an increase in the external magnetic field. At a fixed temperature, an increased external magnetic breaks the magnetic order. This effect can be observed, for example, for $Cr_{0.1}ZrTe_2$, in which the magnetic order is observed at H = 100 Oe (the magnetic susceptibility passes through the maximum), but it is absent at H = 1000 Oe.

**Table 2**. Magnetic characteristics of $Cr_xZrTe_2$ obtained from the temperature dependencies of the magnetic susceptibilities: effective magnetic moment $\mu_{eff}$ per Cr atom, freezing temperature of the magnetic moments $T_c$ for $Cr_xZrTe_2$, measured in weak (100 Oe) and strong (1000 Oe) magnetic fields.

|  | $\mu_{eff}$ ($\mu_B$) 1000 Oe | $T_c$, K 1000 Oe | $T_c$, K 100 Oe |
|---|---|---|---|
| $Cr_{0.05}ZrTe_2$ | 3.8 | n.a. | n.a. |
| $Cr_{0.1}ZrTe_2$ | 3.8 | n.a. | 3.60 |
| $Cr_{0.15}ZrTe_2$ | 4.0 | 4.00 | 6.62 |
| $Cr_{0.2}ZrTe_2$ | 4.1 | 7.20 | 10.11 |
| $Cr_{0.25}ZrTe_2$ | 4.0 | 9.64 | 13.75 |
| $Cr_{0.3}ZrTe_2$ | 3.8 | 9.20 | 14.60 |

Spin-glass (spin-cluster) ordering is typical for the $M_xTiCh_2$ (Ch = S, Se, Te) intercalation compounds (M – 3d-transition metal) [65–68]. However, the influence of the external magnetic field on the ordering temperature and magnetic susceptibility has not been observed previously. Probably, the reason is that the M 3d-transition atoms in all $M_xTiCh_2$ (M = Cr, Fe, Co, and Ni) intercalation compounds occupy the octahedral sites in the interlayer space.

In the studied $Cr_xZrTe_2$ system, the Cr atoms occupy only the octahedral sites at x < 0.1 and at x = 0.25. At other concentrations, the Cr atoms are partially or mainly distributed over the tetrahedral sites. In the case of the distribution of the Cr atoms over the octahedral sites, the distance between the Cr atoms can be any multiple of $a_0$. The $a_0$ lattice parameter varies from the value of 3.4 Å to 3.95 Å for different $M_xTCh_2$ (T = Ti, Zr, Hf; M – 3d-transition metal) [69]. It means that the M intercalated atoms cannot be closer to each other then 3.4 Å. In the case of the Cr atoms occupy the tetrahedral sites in the $Cr_xZrTe_2$ lattice, the minimum possible distance between Cr atoms coincides with the distance between the nearest tetrahedral sites and ranges from 2.508 Å to 2.505 Å. It means that in the case of the Cr atoms occupy the tetrahedral sites, the distance between them can be smaller than that in the case of the Cr atoms occupy the octahedral sites. A decrease in the possible distance between the intercalated atoms obviously leads to an increase in the ferromagnetic contribution to the total magnetic interaction so that it becomes comparable with the antiferromagnetic contribution. This leads to a noticeable effect of the magnetic field on the temperature of the magnetic ordering, providing the field-induced magnetic order-disorder transition.

# Discussion

In general, our results are consistent with those of [57], where the $Cr_{0.4}ZrTe_2$ crystal was studied using ARPES. Indeed, Cr intercalation substantially changes the character of the chemical bonding in the $ZrTe_2$ host lattice. This is indicated by a change in the Zr and Te core level spectra and a significant change in Zr $M_{2,3}$ XAS (see Fig. 3 and Fig. 5). At the same time, a significant (0.16 eV) shift of the Fermi level due to the Cr electrons transfer to the ZrTe$_2$ lattice is observed. Whereas, in [57] a similar shift was not (у Махмуда точно не было сдвига?) observed. This is probably due to the different Cr concentrations in the crystal studied in [57] and in the current work. The composition of the crystal studied in [57] is declared as Cr$_{0.4}$ZrTe$_2$. This concentration is close to the percolation threshold in the sublattice of intercalated atoms (x = 0.5). This, indeed, allows us to expect those effects that were observed in [57]: 1) a strong increase in the binding energy of the valence band since at this Cr concentration each Te atom is coordinated by Cr atom; 2) a significant increase in the Fermi energy due to the charge transfer from the Cr atoms to the ZrTe$_2$ lattice; 3) the formation of the band of the impurity states with a pronounced dispersion (the formation of a dispersionless band of the impurity states due to a weak overlap of the orbitals of intercalated atoms is typical for the M$_x$TCh$_2$ materials with low *x*).

Comparative analysis of the experimental data from [57] and from our study is complicated by the fact that it is not known what type of the crystallographic positions (tetrahedral or octahedral) in the crystal studied in [57] are occupied by the Cr atoms. However, the absence of the superstructure lines in the X-ray diffraction pattern and the high Cr concentration suggest that the Cr atoms are more likely distributed over the tetrahedral sites rather than over the octahedral ones. Therefore, one should expect a weak Cr 3d/Zr 4d hybridization and a strong Cr 3d/Te 4p one for Cr$_{0.4}$ZrTe$_2$. This is indicated by the increase in the binding energy of the Te 4p-derived valence band and a slight change in the binding energy of the Zr 4d-derived conduction band. In the case of Cr$_{0.25}$ZrTe$_2$, the Cr atoms occupy the octahedral sites and the main contribution to the chemical bond between Cr atoms and the host lattice is caused by the Cr 3d/Zr 4d hybridization. The described character of the chemical bonding between the intercalant atoms with the host lattice allows us to control the band gap width in the IV-group LTMDs intercalation compounds choosing the octahedral or tetrahedral coordination of the intercalant atoms by the chalcogen atoms.

An unusual behavior of the magnetic properties of the Cr$_x$ZrTe$_2$ compounds has been revealed from their temperature dependence of the magnetic susceptibility. The temperature of the magnetic ordering decreases with increasing external magnetic field. We associate this effect with the enhancement of the ferromagnetic contribution to the spin-glass magnetic interaction in the Cr sublattice. The observed in [57] drop in the electrical resistance at low temperatures has been explained as a superconducting transition considering the reduce of the transition temperature in the external magnetic field. However, the same drop in the electrical resistance can be caused by the formation of the magnetic order.

The direct overlap of the Cr 3d orbitals in Cr$_x$TiSe$_2$ at x ≥ 0.25 leads to the vanishing of the spin splitting of the Cr 3d states [46]. It was associated with the broadening of the spin subbands followed by their overlap. However, the first principle calculations data for Cr$_x$ZrTe$_2$ show the spin splitting even at x = 0.25. On the other hand, the Cr 3d$^3$ electronic configuration allows a direct overlap of the Cr 3d orbitals [56]. A similar overlap was observed in the ResPES spectra for Fe$_x$TiSe$_2$ [54]. Probably, the spin polarization can be observed only in the case of the Cr 3d orbitals remain not overlapped even at x = 0.25, which corresponds to the percolation threshold in the Zr-Cr-Zr complexes sublattice. It may be caused by the larger value of the $a_0$ lattice parameter for ZrTe$_2$ ($a_0$ = 3.952 Å) as compared to that for TiSe$_2$ ($a_0$ = 3.540 Å). In this case the following question arises: what type of the hybridization stabilizes the observed ordering of the Cr atoms? We can assume that the ordering is caused by the Zr/Te hybridization, but not by the direct overlap of the Cr 3d orbitals. This picture explains the weak effect of the Cr ordering on the value of the effective magnetic moment.

The displacement of the Cr atoms from the octahedral to tetrahedral sites leads to the almost independency of the $c_0$ lattice parameter on the Cr content. This can be caused by the breakup of the Cr 3d/Zr 4d hybridization; as we know, this hybridization lead to the lattice contraction. Therefore, the distribution of the Cr atoms over the octa- and tetrahedral sites can be due to competition between the Coulomb repulsion in the Cr sublattice and the stable state of the ordered Cr atoms in the octahedral sites.

## Conclusions.

The refinement of the crystal structure show that the Cr atoms balance between the occupation of the octa- and tetrahedral sites. The formation of the F2/m superstructure stabilizes the occupation of the octahedral sites. In this case, the ordering of the Cr atoms is accompanied by the Zr 4d/Te 4p hybridization. This ordering along with an increase in the distance between Cr atoms (in comparison with that for the intercalation compounds based on $TiCh_2$, Ch = S, Se, Te) allows to keep the spin splitting of the Cr 3d states near the Fermi level. This suggests that the spin splitting will remain at higher Cr concentrations, which provide higher temperatures of the magnetic ordering.

## Acknowledgments

The research was carried out within the state assignment of Minobrnauki of Russia (theme "Electron" No. AAAA-A18-118020190098-5, theme "Spin" No. AAAA-A18-118020290104-2 and theme "Quantum" No. AAAA-A18-118020190095-4) and with partial financial support of the RFBR (project 20-03-00275). This work has been done using facilities of the Shared Service Centre "Ural-M", Institute of Metallurgy of the Ural Branch of the Russian Academy of Sciences. The sample synthesis were performed within the RSF grant (project No.17-73-10219). I.P., and S.N. acknowledge funding from EUROFEL project (RoadMap Esfri). This project has received funding from the EU-H2020 research and innovation program under grant agreement No 654360 having benefitted from the Access provided by IOM-CNR in Trieste (Italy) within the framework of the NFFA-Europe Transnational Access Activity. We thank Federica Bondino and Elena Magnano for the kind support and we acknowledge the Elettra Sincrotrone Trieste for providing access to its synchrotron radiation facilities.